\documentclass[3p,twocolumn]{elsarticle}
\usepackage{lineno,hyperref,amsmath,amssymb,mathtools,suffix,dutchcal,xcolor}
\newcommand{ \eq}[1]{Eq.~(\ref{eq:#1})}
\newcommand{ \eqq}[2]{Eqs.~(\ref{eq:#1})-(\ref{eq:#2})}
\modulolinenumbers[5]
\biboptions{sort&compress}
\makeatletter
\def\ps@pprintTitle{%
  \let\@oddhead\@empty
  \let\@evenhead\@empty
  \let\@oddfoot\@empty
  \let\@evenfoot\@oddfoot}
\makeatother

\numberwithin{equation}{section}

\def\oo{\infty}
\def\om{\omega}
\def\Li{\textrm{Li}}
\def\hatom{{\hat\om}}
\def\hatt{{\hat t}}
\def\mmu{m_{\mu}}
\def\uuu{v}
\def\arcsin{\textrm{arcsin}}

\def\Re{\mathrm{Re}}

\def\facmm{\frac{m_{\mu}^2}{16\pi^2}}
\newcommand{\gzp}{g_{b}}
\newcommand{\gdi}{g_{5}}
\newcommand{\Fwsm}{F_{02}}
\newcommand{\Fwla}{F_{2\oo}}
\newcommand{\Fwsmpa}{F_{02}^{\textrm{odd}}}
\newcommand{\Fwlapa}{F_{2\oo}^{\textrm{odd}}}

\newcommand{\Fwdi}{F_{5}}
\newcommand{\Fwdipa}{F_{5}^{(1)}}
\newcommand{\Fwdipb}{F_{5}^{(2)}}
\newcommand{\Fwdipc}{F_{5}^{(3)}}
\newcommand{\ii}{\mathrm{i}}

\DeclarePairedDelimiterX\MeijerM[3]{\lparen}{\rparen}%
{\begin{smallmatrix}#1 \\ #2\end{smallmatrix}\delimsize\vert\,#3}

\newcommand\MeijerG[8][]{%
 G^{\,#2,#3}_{#4,#5}\MeijerM[#1]{#6}{#7}{#8}}

\WithSuffix\newcommand\MeijerG*[7]{%
 G^{\,#1,#2}_{#3,#4}\MeijerM*{#5}{#6}{#7}}
\begin{document}
\begin{frontmatter}
\title{{\bf Time-kernel for lattice determinations of NLO hadronic 
\\
   vacuum polarization contributions
   to the muon {\boldmath $g$-2\unboldmath}  }}

\author[mymainaddress,mysecondaryaddress]{Elisa Balzani}
\author[mymainaddress,mysecondaryaddress]{Stefano Laporta}
\author[mysecondaryaddress,mythirdaddress]{Massimo Passera}

\address[mymainaddress]{Dipartimento di Fisica e Astronomia `G.~Galilei', Universit\`a di Padova, Italy}
\address[mysecondaryaddress]{Istituto Nazionale di Fisica Nucleare, Sezione di Padova, Padova, Italy}
\address[mythirdaddress]{Korea Institute for Advanced Study, Seoul, Republic of Korea}

\begin{abstract}
We study the time-momentum
representation
of the kernel needed to compute 
the hadronic vacuum polarization contribution to the muon $g$-$2$ 
in the space-like region at  next-to-leading order. 
For small values of the time, we present  analytical series expansions; 
for large values of the time, we present  numerical series expansions 
which overcome 
the problems showed by na\"ive  asymptotic expansions.
These results are to be employed in lattice QCD determinations of 
the hadronic vacuum polarization contribution to the muon $g$-$2$ 
at  next-to-leading order. 
\end{abstract}
\end{frontmatter}
\section{Introduction}
\label{Introduction}

The Muon $g$-2 (E989) experiment at Fermilab has recently presented 
its second measurement of the muon magnetic moment anomaly,
$a_{\mu} = (g_{\mu}-2)/2$
~\cite{Muong-2:2023cdq}.
It confirms the first one~\cite{Muong-2:2021ojo,Muong-2:2021vma,Muong-2:2021ovs,Muong-2:2021xzz}, 
with a factor of two improvement in precision,
as well as the earlier results of the E821 experiment at 
Brookhaven~\cite{Muong-2:2006rrc}. 
Moreover, in a longer term, also the E34
collaboration at J-PARC~\cite{Abe:2019thb} aims at measuring the muon
$g$-$2$ through a new low-energy approach.

The present muon $g$-2 experimental average shows a
$5 \sigma$ discrepancy with the value of the Standard Model (SM)
$a_{\mu}$ prediction quoted by the Muon $g$-2
Theory Initiative~\cite{Aoyama:2020ynm}.
The main uncertainty of the muon $g$-2 SM prediction
arises 
from its hadronic vacuum polarization (HVP) contribution, 
$a_{\mu}^{\rm HVP}$, which cannot be reliably
computed
perturbatively in QCD and relies on experimental data as input to 
dispersion relations.
Indeed, this contribution has been traditionally 
calculated
via a dispersive, or time-like, integral using 
low--energy $e^{+}e^{-}\to \textrm{hadrons}$ data.
Currently,
the time-like calculation of $a_{\mu}^{\rm HVP}$ includes
the leading-order (LO), next-to-leading-order (NLO)
and next-to-next-to-leading-order (NNLO)
terms~\cite{Jegerlehner:2017gek,Davier:2017zfy,Keshavarzi:2018mgv,Colangelo:2018mtw,Hoferichter:2019mqg,Davier:2019can,Keshavarzi:2019abf,Hoid:2020xjs,Kurz:2014wya}.

An alternative determination of $a_{\mu}^{\rm HVP}$ has been provided
by lattice QCD~\cite{Chakraborty:2017tqp,Borsanyi:2017zdw,Blum:2018mom,Giusti:2019xct,Shintani:2019wai,FermilabLattice:2019ugu,Gerardin:2019rua,Aubin:2019usy,Giusti:2019hkz,Chakraborty:2018iyb}.
In the last few years 
significant progress has been made 
in first-principles lattice QCD calculations of its LO part,
$a_{\mu}^{\rm HVP}({\rm LO})$,
although the precision of these results is,
in general, not yet competitive with that of the time-like
determinations based on experimental data.
In 2021 the BMW collaboration published the first lattice QCD
calculation of $a_{\mu}^{\rm HVP}({\rm LO})$ with an impressive
sub-percent (0.8\%) relative accuracy~\cite{Borsanyi:2020mff}.
This remarkable result weakened the discrepancy between 
the muon $g$-2 SM prediction and the experimentally measured value,
but showed a tension with the time-like data-driven
determinations of $a_{\mu}^{\rm HVP}({\rm LO})$,
being $2.2 \sigma$ higher than the Muon $g$-2 Theory Initiative
data-driven value.
Recently, the BMW result has been improved in Ref.~\cite{Boccaletti:2024guq}.
Moreover, a new measurement of the $e^+ e^- \to \pi^+ \pi^-$
cross section from the CMD-3 experiment disagrees with 
all the other $e^+ e^-$ data~\cite{CMD-3:2023alj,CMD-3:2023rfe}.
Efforts are ongoing to clarify the
current theoretical situation.

A new and competitive determination of $a_{\mu}^{\rm HVP}$
based on a method alternative to the time-like
and lattice QCD ones is therefore desirable.
A new approach to determine the HVP contribution to the
muon $g$-2, measuring the effective electromagnetic coupling
in the space-like region via scattering data,
was proposed in 2015~\cite{CarloniCalame:2015obs}.
The elastic scattering of high-energy muons on atomic electrons
was identified as an ideal process for this measurement,
and a new experiment, MUonE, was proposed at CERN to
measure the shape of the differential cross section
of $\mu$-$e$ elastic scattering as a function of the space-like
squared momentum transfer~\cite{Abbiendi:2016xup,MUonE:LoI,Banerjee:2020tdt}.

In Ref.~\cite{Balzani:2021del}
we investigated the HVP contributions to the muon $g$-2
in the space-like region.
At LO, simple results are long known and form the basis for present
lattice QCD and future MUonE determinations of
$a_{\mu}^{\rm HVP}({\rm LO})$.
In Ref.~\cite{Balzani:2021del}
we  provided 
simple analytical expressions
to extend the space-like calculation of the 
$a_{\mu}^{\rm HVP}$ contribution to NNLO 
(see also Ref.~\cite{Laporta:Strong2021,Passera:Remiddi2021,Nesterenko:2021byp}). 

In principle, the space-like expressions of Ref.~\cite{Balzani:2021del}
can be directly used in lattice
determinations of HVP contributions to the muon $g$-2.
However,
lattice calculations widely use 
the time-momentum representation~\cite{Bernecker:2011gh}.
Therefore, in this paper we will work out the time-momentum representation
of the NLO space-like kernel of Ref.~\cite{Balzani:2021del}
\footnote{These results were presented at the
workshop~\cite{Laporta:Bern2023}}.
\section{The time-kernel at leading order}
\label{HVPLO}
According to Refs.~\cite{Bernecker:2011gh,DellaMorte:2017dyu},
in the time-momentum representation
the LO HVP contribution to the muon $g$-$2$ can be written as 
\begin{equation}
\label{eq:LOtimeintegral}
a_{\mu}^{\rm HVP}(\textrm{LO})=
\left(\frac{\alpha}{\pi}\right)^2
\int_0^{\oo} dt\; G(t)\; \tilde{f}_2(t)
\ ,
\end{equation}
where $t$ is the \emph{Euclidean time}, 
$G(t)$ is the spatially summed two-point correlator of the
electromagnetic current, 
and 
the LO time-kernel 
$\tilde{f}_2(t)$
can be written  as
\begin{equation}
\tilde{f}_2(t)=
8\pi^2 \int_0^{\oo} \frac{d\om}{\om}\  f_2(\om^2) \  g(\om t) \ ,
\end{equation}
where 
\begin{equation}\label{eq:gdefin}
g(w)=w^2-4 \sin^2 \left(\frac{w}{2}\right) \ .
\end{equation}
In the following, we use extensively the adimensional frequency and
time ($m_{\mu}$ is the muon mass)  
\begin{equation}\label{eq:adimdef}
\hatom=\frac{\om}{\mmu} \ , \quad \hatt=\mmu t \; .
\end{equation}
The LO kernel $f_2(\om^2)$ can be written 
as 
\begin{equation}
f_2(\om^2)=
\frac{1}{m_{\mu}^2}
\frac{F_2(1/y(-{\hatom}^2))}{-\hatom^2}\ ,
\end{equation}
where $y(z)$ is the rationalizing variable
\begin{equation}\label{eq:ydef}
         y(z) = \frac{z-\sqrt{z(z-4)}}{z+\sqrt{z(z-4)}},
\end{equation}
and $F_2(y(z))$ is the known LO space-like kernel
written in the form appearing in Ref.~\cite{Balzani:2021del}.
Substituting the expression of $F_2(y)$ from Ref.~\cite{Balzani:2021del}
one obtains
the result
\begin{equation}
f_2(\om^2)=
\frac{1}{m_{\mu}^2}
\frac{1}{y(-\hatom^2)(1-y^2(-\hatom^2))}\ .
\end{equation}

The integration over $\om$ is complicated~\cite{DellaMorte:2017dyu},
the result
contains a Meijer $G$-function:
\begin{align}\label{eq:timekernelLO}
\frac{m_{\mu}^2}{8\pi^2}
\tilde{f}_2(t)&=
\tfrac{1}{4}
\MeijerG*{2}{1}{1}{3}{\frac{3}{2}}{0,1,\frac{1}{2}}{{\hatt}^2}
+\tfrac{{\hatt}^2}{4}
+\tfrac{1}{{\hatt}^2}
\cr &
+2 \ln ({\hatt})
-\tfrac{2}{{\hatt}} K_1(2 {\hatt})
+2 \gamma
 -\tfrac{1}{2}\ ,
\end{align}
where $K_n$ is the modified Bessel function of the second kind.
This expression can also be written in terms of integrals
of the Bessel functions instead of the Meijer $G$-function (see a similar integral in
Ref.~\cite{Frolov:2011rm}). 
This can be done by applying 
the identity
\begin{multline}
\label{eq:mejertostruve}
 \MeijerG*{2}{1}{1}{3}{\frac{3}{2}}{0,1,\frac{1}{2}}{u^2}=
-2
+8 \int^u_0 dv\  (v-u) K_0(2v)
\\
= -4u\biggl[
 K_1(2u)
 + \pi u
\biggl(
K_0(2u)\mathbf{L}_{-1}(2u)
\\
+K_1(2u)\mathbf{L}_{0}(2u)
\biggr)
\biggr]
\end{multline}
to~\eq{timekernelLO},
where $\mathbf{L}_{-1}$ and $\mathbf{L}_0$ are modified Struve functions.
\section{The time-kernel at NLO } 
\begin{figure}
\begin{center}
\includegraphics[width=\columnwidth]{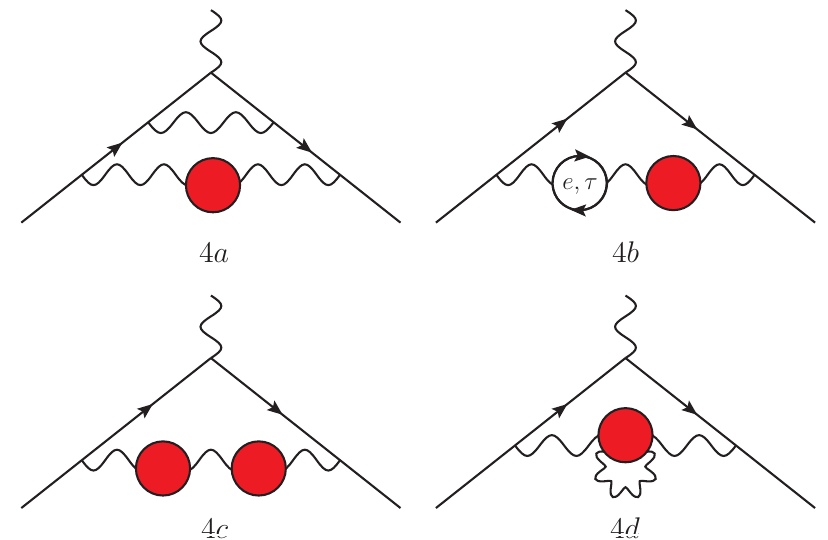}
\caption{Sample $\mathcal{O}(\alpha^3)$ diagrams contributing to the HVP corrections to the muon $g$-2.}
\label{fig:NLOdiagrams}
\end{center}
\end{figure}

\color{black}
The HVP contribution to the muon $g$-2 at NLO, $a_{\mu}^{\rm HVP}({\rm NLO})$, 
is due to $\mathcal{O}(\alpha^3)$ diagrams that can be classified as
discussed in Ref.~\cite{Balzani:2021del} (see Fig.~\ref{fig:NLOdiagrams}).
%
%

We will focus here on class $(4a)$,
by far the most challenging one,
which also yields the largest numerical contribution.
The methods presented in this letter can also be applied to classes
$(4b)$ and $(4c)$, see Ref.~\cite{Wittig2024:4bc}.
Class $(4d)$ is not considered 
for  $a_{\mu}^{\rm HVP}({\rm NLO})$ 
because its
contribution is already  incorporated into $a_{\mu}^{\rm HVP}({\rm LO})$. 


\color{black}
Similarly to the LO case, 
the NLO contribution
of class $(4a)$
can be written as
\begin{equation}
\label{eq:NLO4atimeintegral}
a_{\mu}^{\rm HVP}(\textrm{NLO};4a)=
\left(\frac{\alpha}{\pi}\right)^3
\int_0^{\oo} dt \;G(t)\; \tilde{f}_4(t) 
\ ,
\end{equation}
where $\tilde{f}_4(t)$ is the NLO time-kernel
\begin{equation}\label{eq:intk4}
\tilde{f}_4(t)=
8\pi^2 \int_0^{\oo} \frac{d\om}{\om}\ f_4(\om^2)\  g(\om t)\ .
\end{equation}
For convenience we define an adimensional $\hat{f}_4(\hatom^2)$:
\begin{equation}
f_4(\om^2)=\frac{1}{m_{\mu}^2} \hat{f}_4(\hatom^2) \ .
\end{equation}
$\hat{f}_4(\hatom^2)$ is related to the NLO space-like kernel $F_4(y)$ 
obtained in Ref.~\cite{Balzani:2021del}
\begin{equation}
\label{eq:f4F4}
\hat{f}_4(\hatom^2)=
\frac{2\; F_4(1/y(-\hatom^2))}{-\hatom^2}\ .
\end{equation}

By splitting $g(w)$,
the integral~(\ref{eq:intk4}) is split into  two parts:
\begin{equation}\label{eq:f4ab}
\tilde{f}_4(t)=
\tilde{f}_4^{(a)}(t)
+\tilde{f}_4^{(b)}(t)\ .
\end{equation}
The first part is 
\begin{equation}\label{eq:f4a}
\frac{m_{\mu}^2}{8\pi^2}
\tilde{f}_4^{(a)}(t)=
\int_0^{\oo} \frac{d\hatom}{\hatom}\; {\hat{f}_4(\hatom^2)}
\left( \hatom^2 \hatt^2 \right) \ .
\end{equation}
The integral in~\eq{f4a} can be calculated in analytical form;
in fact, 
substituting the expression~\eq{f4F4}, 
performing the change
of variable $\hatom \to y $, recalling 
from Ref.~\cite{Balzani:2021del}
that $F_4(z)$ is the imaginary part of
the timelike kernel $K_4(z)$, and using the dispersive integral for $K_4(z)$ 
one obtains 
\begin{align}\label{eq:f4a2}
\facmm
&
\tilde{f}_4^{(a)}(t)=
\tfrac{\hatt^2}{2}  \int_{-\oo}^0 \frac{dz}{z}\;F_4(1/y(z))=
\cr&
\tfrac{\hatt^2}{2} \int_{-\oo}^0 \frac{dz}{z}\;\frac{1}{\pi} {{\rm Im} K_4(z)}=
\tfrac{\hatt^2}{2} K_4(0)=
\cr&
\tfrac{\hatt^2}{2} \biggl(
\tfrac{197}{144}
+\tfrac{\pi ^2}{12}
-\tfrac{1}{2}\pi ^2 \ln 2
+\tfrac{3}{4} \zeta(3)
\biggr)
\;.
\end{align}
In the above expression $K_4(0)$ is the value of the 2-loop $g$-$2$,
and we have also incorporated the factor 2 from~\eq{f4F4} in the
denominator $16 \pi^2$.

The second part
of~\eq{f4ab}
is  
\begin{equation}\label{eq:f4b}
\frac{m_{\mu}^2}{8\pi^2}
\tilde{f}_4^{(b)}(t)
=
\int_0^{\oo} 
\frac{d\hatom}{\hatom}
\hat{f}_4(\hatom^2)
\left(
-4
\sin^2 \left(\tfrac{\hatom \hatt}{2}\right)
\right)\ .
\end{equation}
Substituting the expression (\ref{eq:f4F4}) in~\eq{f4b},
one finds that
the integrand contains 
$\ln y$, $\ln(1\pm y)$, $\Li_2(\pm y)$.
The integration of single logarithms and product of logarithms 
can be done analitically, obtaining complicated expressions 
containing several Bessel functions,
exponential integrals, Meijer $G$-functions. 
But, unfortunately, we were not able to calculate analytically the integrals
containing the dilogarithms of $y$.

As an alternative, in the next sections  we will work out some  
series expansions of $\tilde{f}_4(t)$.

\section{Expansion for small $t$}
In this section we work out the expansion of the NLO time-kernel
(\ref{eq:intk4}) for
$\hatt \ll 1 $.
We split the interval of integration at an intermediate point
$ \hatom_0(\hatt)$:
\begin{align}
\label{eq:splitom}
\int_0^{\oo}             \frac{d\hatom}{\hatom}\;\hat{f}_4(\hatom^2) g(\hatom \hatt) =&
\int_0^{\hatom_0(\hatt)} \frac{d\hatom}{\hatom}\;\hat{f}_4(\hatom^2) g(\hatom \hatt)
\cr
+&
\int_{\hatom_0(\hatt)}^{\oo}
                         \frac{d\hatom}{\hatom}\;\hat{f}_4(\hatom^2) g(\hatom \hatt)\ .
\end{align}
The value of the integral is independent of the point of splitting
$\hatom_0$; a convenient choice is 
\begin{equation}
\hatom_0(\hatt)=\frac{1-\hatt}{\sqrt{\hatt}} \gg 1 \ .
\end{equation}
In the integral 
over the interval $ [0,{\hatom_0(\hatt)}]$ of~\eq{splitom},
first we expand in series $g(\hatom \hatt)$ 
for $ \hatt \ll 1$,
then 
we make the convenient change of variable $\hatom \to y=y(-\hatom^2)$
(see~\eq{f4F4}),
and 
we integrate  over the interval 
$-1/\hatt=y\left(-\hatom_0^2\right) \le y \le -1 $.
In the second integral 
over the interval $ [\hatom_0(\hatt),\oo) $, first
we expand in series $\hat{f}_4(\hatom^2)$ for $\hatom \gg 1
$, then we integrate  over $\hatom$.
The whole ${\tilde f}^{(4)}(t)$ is obtained summing up
the results of the two integrals.
The expansion turns out to have the form 
\begin{multline}
\label{eq:f4smallt}
\facmm \tilde{f}_4(t)=
\sum_{\substack{n \ge 4 \\ n \ \textrm{even} } } 
\frac{\hatt^{n}}{{n}!} 
\biggl[
a_n
+b_n  \pi^2
+c_n \left(\ln(\hatt)+\gamma\right)
\\
+d_n \left(\ln(\hatt)+\gamma\right)^2
\biggr]\; ;
\end{multline}
The analytical values of the coefficients $a_n$, $b_n$, $c_n$, $d_n$ of the expansion 
up to $\hatt^{30}$ are available in Table~\ref{table:fsmallt}.

\section{Asymptotic expansions for large $t$}
We decompose  ${\tilde f}_4^{(b)}(t)$ 
in two parts, according to the different behaviour for $t \to \oo$,
\begin{equation}
   {\tilde f}_4^{(b)}(t)= 
   {\tilde f}_4^{(b;1)}(t) 
 + {\tilde f}_4^{(b;2)}(t) \,.
\end{equation}
\subsection{Main contribution}
${\tilde f}_4^{(b;1)}(t)$ is the dominant contribution, and its
asymptotic expansion contains powers of $1/\hatt$:  
\begin{equation}\label{eq:f4b1asym}
{\tilde f}_4^{(b;1)}(t) = A_0+A_1 \hatt + B_0 \ln \hatt 
+ B_2 \frac{\ln \hatt}{\hatt^2} +\sum_{n=1}^{\oo}\frac{C_n}{\hatt^n} \ . 
\end{equation}
The integral representation of 
${\tilde f}_4^{(b;1)}(t)$ can be obtained in this way:
first, we split the integrand of~\eq{f4b},
\begin{align}\label{eq:f4b1rew}
\frac{m_{\mu}^2}{8\pi^2} \tilde{f}_4^{(b;1)}(t)
=&
-2 \lim_{\epsilon \to 0}
\biggl[
 \int^{\oo}_{\epsilon} \frac{d\hatom}{\hatom} \hat{f}_4(\hatom^2) 
\nonumber \\
&- \int^{\oo}_{\epsilon} \frac{d\hatom}{\hatom} \hat{f}_4(\hatom^2)
\cos \left(\hatom \hatt\right)
\biggr]
\,, 
\end{align}
where $\epsilon$ regulates the divergence in $\hatom=0$;
subsequently, we expand in series $\hat{f}_4(\hatom^2)$ around $\hatom=0$ 
in  the second integral of~\eq{f4b1rew},
\begin{equation}
\hat{f}_4(\hatom^2)=
\frac{1}{8\hatom}
-\frac{1}{2}
+
\left(\frac{\ln \hatom }{2}
   +\frac{251}{2880}
   +\ln 2\right)\hatom 
+{\ldots} \ .
\end{equation}
Formally integrating term by term over $\hatom$ 
using  
\begin{equation}
 \int d\hatom \;\hatom^n \cos(\hatom\; \hatt) = - n! \sin\left(n \pi/2\right)
 \hatt^{-1-n}  \ ,
\end{equation}
we can obtain
the coefficients $A_n$, $B_n$, $C_n$ of~\eq{f4b1asym}.

In section~\ref{sec:furthsub} we will need the first terms of the
asymptotic expansion: 
\begin{multline}
\label{eq:b1asy}
\facmm
\tilde{f}_4^{(b;1)}(t)=
-\frac{\pi  \hatt}{8}
+\ln \hatt
-\frac{7 \zeta(3)}{4}
+\frac{7}{6} \pi ^2 \ln (2)
\\
-\frac{127 \pi ^2}{144}
+\gamma
+\frac{653}{216}
-\frac{5 \left(\ln \hatt +\gamma \right)}{12 \hatt^2}
-\frac{\pi }{2 \hatt}
+\frac{209}{180 \hatt^2}
+\frac{277 \pi }{360 \hatt^3}
\\
+O\left(\frac{1}{\hatt^{4}}\right)
\ .
\end{multline}

\subsection{Exponentially suppressed contribution}
${\tilde f}_4^{(b;2)}(t)$  is the exponentially suppressed contribution.
Its asymptotic expansion contains the 
factor $e^{-2\hatt}$:
\begin{equation}\label{eq:f4b2asym}
{\tilde f}_4^{(b;2)}(t) = e^{-2\hatt} 
\sum_{n=0}^{\oo} 
\left( D_n+ \frac{E_n \ln \hatt+F_n}{\sqrt{\hatt}} \right)
\frac{1}{\hatt^n} \ ,
\end{equation}
where $D_n$, $E_n$ and $F_n$ are constants.

${\tilde f}_4^{(b;2)}(t)$  has also a representation as  an integral 
over the contour $\cal{C}$ shown in fig.\ref{fig:pathom}:
\begin{figure}
\begin{center}
\includegraphics[width=0.8\columnwidth]{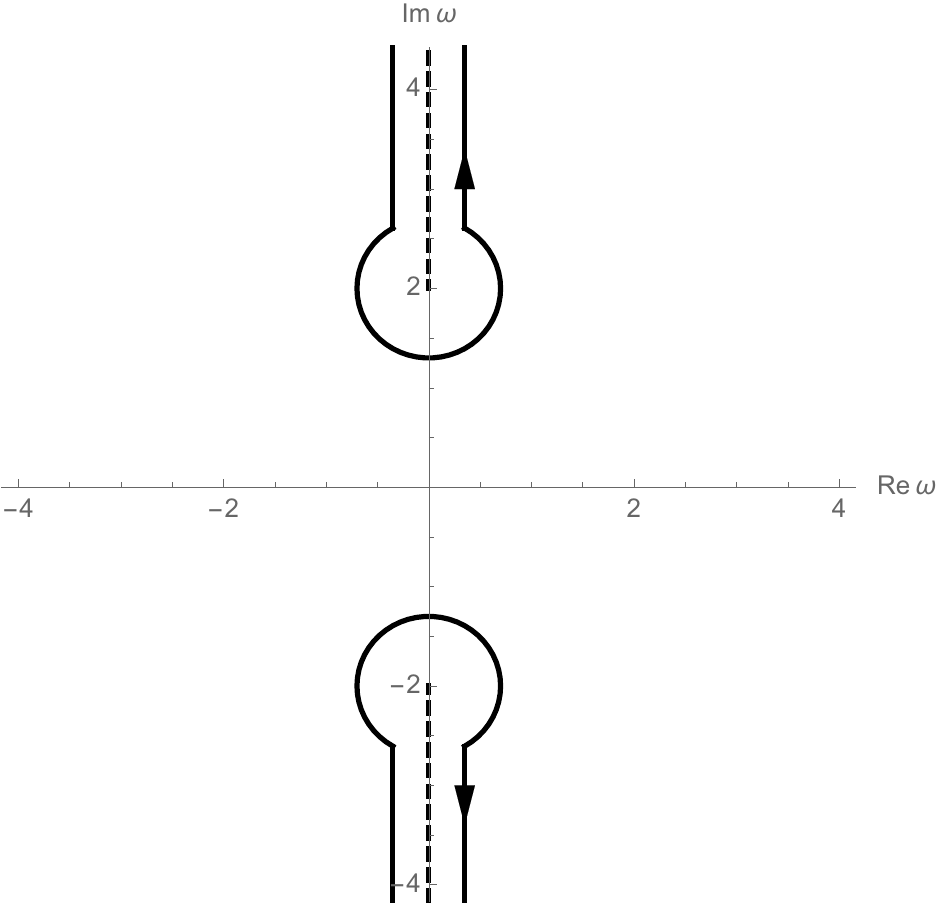}
\caption{The thick lines show the path of integration around the
discontinuities (the dashed lines) giving the exponentially suppressed contribution.}
\label{fig:pathom}
\end{center}
\end{figure}

\begin{equation}
\label{eq:f4b2C}
\frac{m_{\mu}^2}{8\pi^2}
{\tilde{f}_4^{(b;2)}(t)}=
\int_{\cal C} \frac{d\hatom}{\hatom} \hat{f}_4(\hatom^2) 
2 \cos \left(\hatom \hatt\right) \ .
\end{equation}
The presence of the exponential factor is 
due to the singularities of the integrand in $\hatom=\pm 2i$,
which come from  terms of $\hat{f}_4(\hatom)$ 
containing $\sqrt{\hatom^2+4}$. 

Due to their asymptotic nature,
the 
expansions (\ref{eq:f4b1asym}) and  (\ref{eq:f4b2asym})
have a limited use for numerical evaluations. 
Increasing $n$, the coefficients 
grow factorially and, therefore,  
one needs to truncate the series at the index $n=\bar n(\hatt)$
where the terms start increasing.
One can
show that the error due to the truncation of the first series is of
the same order of magnitude of the  value of the second series,
making the inclusion  of the exponentially suppressed  contribution 
meaningless.

Due to this fact,
in the next sections we will explore a different approach, 
able to find expansions around a \emph{finite}
point $\hatt=\hatt_0$  converging for $\hatt \to \oo$. 

\section{$w$-Integral representation for $\tilde{f}_4^{(b)}(t)$}
We start from the definition (\ref{eq:f4b}) of
 $\tilde{f}_4^{(b)}(t)$.
In the integrand we add and subtract
to $\tfrac{\hat{f}_4(\hatom^2)}{\hatom}$
a piece $h_0(\hatom)$ 
which contains the terms non
integrable in $\hatom=0$ or $\hatom=\pm 2i $,
\begin{equation}
h_0(\hatom)=
 \frac {1}{8\hatom^2}
+\frac{\hatom \pi}{16 (4 + \hatom^2)^{3/2}} - \frac{\pi}{2 (4 +
\hatom^2)} \ .
\end{equation}
Defining
\begin{equation}
\gzp(\hatom)=\frac{\hat{f}_4(\hatom^2)}{\hatom}-h_0(\hatom) \\
\end{equation}
and
\begin{align}
\tilde{h}_0(\hatt)&=\
\int_0^{\oo} 2\left( \cos(\hatom \hatt )-1\right) h_0(\hatom)\; d\hatom 
\\
&=\frac{\pi  \hatt}{16}
+ \frac{\pi^2}{8}\left( e^{-2 \hatt} -1 \right)
\cr & 
\phantom{=}
+ \frac{1}{32} \pi^2 \hatt \left( K_0(2 \hatt) -\mathbf{L}_0(2 \hatt)
\right)
\ ,
\end{align}
we write
\begin{align}
\facmm
\tilde{f}_4^{(b)}(t)=
\tilde{h}_0(\hatt)
+\int_0^{\oo} d\hatom\; 2 \left(\cos(\hatom \hatt )-1\right)
\gzp(\hatom)
\ .
\end{align}

Let us consider only the integral with the cosine.
We decompose the cosine in exponentials 
\begin{align}
\int_0^{\oo}d\hatom \; \gzp(\hatom) \cos{\hatom \hatt} 
=\int_0^{\oo}d\hatom \;\gzp(\hatom) \frac{e^{\ii \hatom \hatt} +e^{-\ii
\hatom \hatt}}{2} \;
\ .
\end{align}
Now we split the integral, rotate of $\pi/2$ the
integration path in the complex-$\hatom$ plane and make the change of variable 
$\hatom \to i w $ in the first exponential; then  we rotate of $-\pi/2$ and
make the change
$\hatom \to -i w $
in the second one
\begin{multline}
 \int_0^{ \ii \oo}d\hatom \;\gzp(\hatom) \frac{e^{ \ii \hatom \hatt}}{2} 
+\int_0^{-\ii \oo}d\hatom \;\gzp(\hatom) \frac{e^{-\ii \hatom \hatt}}{2} =
\\
\int_0^{\oo}dw\; F_{0\oo}(w) e^{-w \hatt} \ ,
\end{multline}
where $F_{0\oo}$ is 
\begin{equation}
F_{0\oo}(w)=\frac{\ii}{2} \lim_{\epsilon \to 0^{+}}
\left(
 \gzp(\epsilon+\ii w)  
-\gzp(\epsilon-\ii w)  
\right) \ .
\end{equation}
We have introduced the regulator $\epsilon$ to make sure 
that the integration path remains in the half-plane $\Re(\hatom)>0$.

Due to the presence of the discontinuity, the limit is different
if $0<w<2$ or $w>2$
\begin{equation}
F_{0\oo}(w) = \left \{
\begin{aligned}
      &\Fwsm(w), && \text{if}\ 0<w<2 \ ,\\
      &\Fwla(w), && \text{if}\ w>2   \ .
\end{aligned} 
\right.
\end{equation}

Finally the total integral becomes 
\begin{multline}\label{eq:intwf4b}
\facmm
\tilde{f}_4^{(b)}(t)=
\tilde{h}_0(\hatt) 
+\int_0^2 dw \; \Fwsm(w) 2 \left(e^{-w \hatt}-1\right)
\\
+\int_2^{\oo} dw \; \Fwla(w) 2 \left(e^{-w \hatt}-1\right) \ ,
\end{multline}
where
\begin{multline}\label{eq:f02}
\Fwsm(w)=
\frac{4}{3 w^3}+\frac{w}{16 \left(w^2-4\right)}\\
+\pi  \sqrt{4-w^2}
   \left(\frac{w}{16 \left(w^2-4\right)^2}-\frac{1}{8
      w^2}+\frac{7}{48}\right) \\
      +\biggl[
      \sqrt{4-w^2} \biggl(-\frac{4}{3 w^4}-\frac{17}{48 w^2}
      -\frac{5}{16 \left(w^2-4\right)} \\
      -\frac{1}{4  \left(w^2-4\right)^2}+\frac{1}{8}\biggr) 
   +\pi  \left(\frac{1}{2 w^3}+\frac{w}{2}-\frac{7}{6 w}\right)
   \biggl]\times\\
   \arcsin \left(\frac{w}{2}\right) 
    +\frac{23 w}{144}-\frac{37}{144 w}+\frac{5}{24} w \ln (w)
    \ ,
\end{multline}
\begin{multline}\label{eq:f2oo}
\Fwla(w)=
\frac{4}{3 w^3}+\frac{w}{16\left(w^2-4\right)} \\
+\left(\frac{7}{24}-\frac{1}{4 w^2}\right) \sqrt{w^2-4}
\ln \left(w \left(w^2-4\right)\right) \\
+\sqrt{w^2-4}
      \left(-\frac{1}{3 w^4}+\frac{115}{144 w^2}+\frac{23}{144
         \left(w^2-4\right)}-\frac{23}{144}\right) \\
 +\biggl[-\frac{4}{3 w^5}+\frac{7}{6 w^3}+\frac{w}{2 \left(w^2-4\right)}
  -\frac{29  w}{24}+\frac{47}{12 w} \\
-\sqrt{w^2-4} 
  \biggl(-\frac{4}{3 w^4}-\frac{17}{48 w^2}
  -\frac{5}{16 \left(w^2-4\right)}\\
  -\frac{1}{4 \left(w^2-4\right)^2}
  +\frac{1}{8}\biggr)
    \biggr]
    \frac{\ln (y(w^2))}{2}
    +\frac{23 w}{144}
    -\frac{37}{144 w}
    \\
    +\frac{5}{24} w \ln (w)
 - \left(\frac{1}{w^3}+w-\frac{7}{3 w}\right) 
 L(y(w^2)) \ ,
\end{multline}
\begin{multline}
L(x)=\Li_2(-x)+2 \Li_2(x) \\
 +\tfrac{1}{2} \ln x
\left(\ln(1+x)+2\ln(1-x)\right)\ ;
\end{multline}
$y(z)$ was defined in~\eq{ydef}.

We can also integrate analytically over $w$ the terms of~\eq{intwf4b}
not containing the exponential,
but we have to add and subtract the pole term of the Laurent expansion of
$\Fwsm(w)$ in $w=0$
\begin{equation}
\Fwsm(w)=-\frac{1}{2w}+O(1)\ ,
\end{equation}
obtaining
\begin{align}\label{eq:intwf4bmod}
\facmm
\tilde{f}_4^{(b)}(t)&=
c_0
+\tilde{h}_0(\hatt) 
+\tilde{h}_3(\hatt) 
\nonumber \\
&+\int_0^2 dw \; 2\left( \Fwsm(w)+\frac{1}{2w}\right)  e^{-w \hatt}
\nonumber \\
&+\int_2^{\oo} dw \; 2 \Fwla(w) \; e^{-w \hatt}
\ ,
\end{align}
where
\begin{align}
 c_0=&-2\int_0^2 dw \; \left(\Fwsm(w)+\frac{1}{2w}\right) 
 \nonumber \\
&-2\int_2^{\oo} dw \; \Fwla(w) =
 \frac{653}{216}
+\frac{\pi }{16} 
-\ln(2) 
\nonumber \\
&-\frac{163 }{144}\pi^2
+\frac{7}{6} \pi ^2 \ln (2)
-\frac{7 \zeta (3)}{4}
\end{align}
and
\begin{align}
\tilde{h}_3(\hatt) &=
\int_0^2 dw\; \frac{1-e^{-w \hatt }}{w}
\nonumber \\
&= -\text{Ei}(-2 \hatt)+\ln (2 \hatt)+\gamma
\ .
\end{align}

\section{$w$-integral for exponentially suppressed contribution
$\tilde{f}_4^{(b;2)}(t)$ 
}
We proceed similarly to the previous section.
In~\eq{f4b2C} 
we  add and subtract the pole term $h_2(\hatom)$ of the Laurent expansion of
$\hat{f}_4(\hatom^2)/\hatom$ in $\hatom= \pm 2i$, obtaining 
\begin{align}
\label{eq:wintexpsup1}
\facmm
\tilde{f}_4^{(b;2)}&(t)=
\tilde{h}_2(\hatt) +
\int_{\cal C}d\hatom\;\gdi(\hatom) 2\cos(\hatom \hatt) \ , \\
\gdi(\hatom)&=\frac{\hat{f}_4(\hatom^2)}{\hatom}-h_2(\hatom) \;,\\
h_2(\hatom)&=
- \frac{\pi}{2 (4 + \hatom^2)} \;, \\
\tilde{h}_2(\hatt)&=
\int_0^{\oo} d\hatom \; 2 \cos (\hatom \hatt ) h_2(\hatom) = -\frac{\pi^2}{4}
e^{-2 \hatt} \ .
\end{align}
We consider a path $\cal C$ infinitesimally near the cuts
(see  Fig.\ref{fig:pathom}); 
we decompose the cosine  and make the suitable change of variables in
order to parametrize the two parts of $\cal{C}$ with the same $w$.
We also have to take 
the difference between the values of $\gdi$
between the two cuts, and on the left and the right of  each cut:
\begin{multline}\label{eq:f5}
\Fwdi(w)=\frac{\ii}{2} 
\biggl[
 \lim_{\epsilon \to 0^{+}}  \gdi(\epsilon+\ii w)  
-\lim_{\epsilon \to 0^{-}}  \gdi(\epsilon+\ii w)  
\\
-\lim_{\epsilon \to 0^{+}}  \gdi(\epsilon-\ii w)  
+\lim_{\epsilon \to 0^{-}}  \gdi(\epsilon-\ii w)  
\biggr]\;. 
\end{multline}
Finally 
\begin{equation}
\label{eq:intwf4b2}
\facmm
\tilde{f}_4^{(b;2)}(t)=
\tilde{h}_2(\hatt) +
\int_2^{\oo}dw\; \Fwdi(w) 2 e^{-w \hatt} \;,
\end{equation}
where
\begin{multline}
\Fwdi(w)=
\frac{-23 w^6+230 w^4-508 w^2+192}{144 w^4 \sqrt{w^2-4}}\\
-\frac{-29 w^8+222 w^6-348 w^4-144 w^2+128}{48 w^5 \left(w^2-4\right)}
\ln (y(w^2)) 
\\
-
\left(\frac{1}{w^3}+w-\frac{7}{3 w}\right)
\biggl(
 L(y(w^2)) + \frac{\pi ^2}{4}
\biggr)
\\
+\left(
\frac{7}{24}-\frac{1}{4 w^2}
\right)
\sqrt{w^2-4} \ln \left(w(w^2-4)\right)
\ .
\end{multline}
We note that  the asymptotic expansion ~\eq{f4b2asym} 
could be obtained from the integral representation of~\eq{intwf4b2}, 
by expanding
$\Fwdi(w)$ and $e^{-w \hatt}$ in $w=2$ and by integrating term-by-term over $w$.  
The expansion of $e^{-w \hatt}$ generates the exponential factor
$e^{-2\hatt}$.

We also note that $\Fwla(w)$ also generates all the exponentially suppressed
contributions generated by $\Fwdi(w)$; 
in fact 
we can check that the difference 
$\Fwla(w)-\Fwdi(w)$ is a function regular in $w=2$
\footnote{
Not all the parts of $\hat{f}_4(\hatom^2)/\hatom$ which have
a discontinuity for $\hatom^2 < -4$, once integrated over $\hatom$ give
terms whose asymptotic behaviour contains $e^{-2 \hatt}$ terms. 
An example comes from the second term of $h_0(\hatom)$: its asymptotic
expansion
$
\int_0^{\oo} d\hatom\; 
\tfrac{2 \hatom\cos(\hatom \hatt )}{(\hatom^2+4)^{3/2}} = 
-\tfrac{1}{4 \hatt^2}-\tfrac{9}{16 \hatt^4} -\tfrac{225}{64 \hatt^6}
+O\left(\tfrac{1}{\hatt^8}\right)
$
does not contain $e^{-2 \hatt}$
}.
\section{Further subdivisions of $\tilde{f}_4^{(b)}(t)$ }
\label{sec:furthsub}
Now we have $w$-integral representations: 
\eq{intwf4bmod} for 
$\tilde{f}_4^{(b)}(t)=\tilde{f}_4^{(b;1)}(t)+\tilde{f}_4^{(b;2)}(t)$
and~\eq{intwf4b2}  for $\tilde{f}_4^{(b;2)}(t)$.
In~\eq{f4b1asym} and~\eq{f4b2asym} we have shown also the
general form of their asymptotic expansions. 
Each of these expansions contains contributions with slightly
different behaviour.
In order to obtain numerically efficient expansions around
 finite $\hatt$, we have to introduce further splitting,
separating even and odd powers in $\tilde{f}_4^{(b;1)}(t)$ and
integer and half-integer powers, and logarithms in  $\tilde{f}_4^{(b;2)}(t)$. 

Therefore we subdivide 
$\tilde{f}_4^{(b;1)}(t)$  and 
$\tilde{f}_4^{(b;2)}(t)$ 
in 3 parts, according to their asymptotic behaviour: 
\begin{align}
\tilde{f}_4^{(b;1)}(t)&=
 \tilde{f}_4^{(b;1;1)}(t)
+\tilde{f}_4^{(b;1;2)}(t)
+\tilde{f}_4^{(b;1;3)}(t) 
\ ,
\cr 
\tilde{f}_4^{(b;2)}(t)&=
 \tilde{f}_4^{(b;2;1)}(t)
+\tilde{f}_4^{(b;2;2)}(t)
+\tilde{f}_4^{(b;2;3)}(t)
\ ,
\end{align}
where 
\begin{align}
\label{eq:b11}
\facmm
\tilde{f}_4^{(b;1;1)}(t)&\sim\frac{1}{\hatt}+O\left(\frac{1}{\hatt^3}\right), \\
\label{eq:b12}
\facmm
\tilde{f}_4^{(b;1;2)}(t)&\sim\frac{1}{\hatt^2}+O\left(\frac{1}{\hatt^4}\right), \\
\label{eq:b21}
\facmm
\tilde{f}_4^{(b;2;1)}(t)&\sim
{e^{-2\hatt}}\left[1+O\left(\frac{1}{\hatt^2}\right) \right], \\
\label{eq:b22}
\facmm
\tilde{f}_4^{(b;2;2)}(t)&\sim
{e^{-2\hatt}}\frac{\ln(\hatt)}{\sqrt{\hatt}} \left[1+O\left(\frac{1}{\hatt}\right) \right],\\
\label{eq:b23}
\facmm
\tilde{f}_4^{(b;2;3)}(t)&\sim
{e^{-2\hatt}}\frac{1}{\sqrt{\hatt}}
\left[1+O\left(\frac{1}{\hatt}\right) \right]\ ,
\end{align}
and
$\tilde{f}_4^{(b;1;3)}(t)$ contains the part not included in the above asymptotic
expansions:
\begin{multline}
\label{eq:b13}
\facmm
\tilde{f}_4^{(b;1;3)}(t)=
  \frac{653}{216}
 -\frac{127 \pi ^2}{144}
 -\frac{7 \zeta (3)}{4}
 +\frac{7}{6} \pi ^2 \ln (2) \\
+\left(\ln \hatt +\gamma\right) \left( 1 -\frac{5}{12 \hatt^2}\right)
-\frac{\pi \hatt}{8}\ .
\end{multline} 

\subsection{Subdivision of the exponentially suppressed contribution}
By analizing the asymptotic of each term of $\Fwdi(w)$, it is possible
to separate the contributions to the separate parts of
$\tilde{f}_4^{(b;2)}(t)$, \eq{b21}, \eq{b22}, \eq{b23}.

We find
\begin{align}
\label{eq:wintb21}
\facmm \tilde{f}_4^{(b;2;1)}(t)&=
\tilde{h}_2(\hatt)+\int_2^{\oo}dw\; 2\Fwdipa(w) e^{-w \hatt}
\;, \\
\label{eq:wintb22}
\facmm \tilde{f}_4^{(b;2;2)}(t)&=
\ln \hatt
\int_2^{\oo}dw\; 2\Fwdipb(w) e^{-w \hatt}
\;, \\
\label{eq:wintb23}
\facmm \tilde{f}_4^{(b;2;3)}(t)&=
\int_2^{\oo}dw\; 2 \Fwdipc(w) e^{-w \hatt}
\;, 
\end{align}
where
\begin{equation}
\label{eq:wexpart1}
\Fwdipa(w)=
\frac{\pi^2}{4}\left(
\frac{7}{3 w} -w -\frac{1}{w^3}
\right)\; ,
\end{equation}
\begin{multline}
\label{eq:wexpart2}
\Fwdipb(w)=
\frac{1}{2}\biggl(\sqrt{w^2-4}
\left(\frac{1}{4 w^2}-\frac{7}{24}\right)
\\
-\frac{1}{2}\left(
\frac{1}{w^3}
+ w-\frac{7}{3 w}\right) 
\ln (y(w^2)) 
\biggr)\; ,
\end{multline}
\begin{equation}
\label{eq:wexpart3}
\Fwdipc(w)=
 \Fwdi(w) 
-\Fwdipa(w)
-\Fwdipb(w)\ln \hatt\; .
\end{equation}

\subsection{
Subdivision of the main asymptotic contribution}
We can separate the parts of 
$\Fwsm(w)$ and $\Fwla(w)$
which  generate 
the odd and the even powers 
of $1/\hatt$,
$\tilde{f}_4^{(b;1;1)}(t)$ and
$\tilde{f}_4^{(b;1;2)}(t)$.
Note that the odd powers have a factor $\pi$,
see~\eq{b1asy}:
\begin{align}
\label{eq:wintb11}
\facmm
\tilde{f}_4^{(b;1;1)}(t) =&
\int_0^{2}dw\; 2\Fwsmpa(w) e^{-w \hatt}
\nonumber \\
+&\int_2^{\oo}dw\; 2\Fwlapa(w) e^{-w \hatt}
\;,
\end{align}
where
\begin{align}
\Fwsmpa(w)&=\frac{\pi}{2}\biggl(
\sqrt{4-w^2}
   \left(\frac{7}{24}-\frac{1}{4 w^2}\right)
\cr
& \quad
+
\left(\frac{1}{w^3}+w-\frac{7}{3 w}\right)
\arcsin \left(\frac{w}{2}\right)
\biggr)
\;,
\cr
\Fwlapa(w)&=
\frac{\pi^2}{4}  \left(\frac{1}{w^3}+w-\frac{7}{3 w}\right)
\; .
\end{align} 
The part with even powers of $1/\hatt$ can be found subtracting
everything from the whole integral
\begin{align}
\label{eq:wintb12}
&\facmm 
\tilde{f}_4^{(b;1;2)}(t)=
\nonumber \\
&~~~c_0
-\hat{f}_4^{(b;1;3)}  (t)
-\tilde{h}_2(\hatt) 
+\tilde{h}_0(\hatt) 
+\tilde{h}_3(\hatt) 
\nonumber \\
&~~~+\int_0^2 dw \; 2\left( \Fwsm(w)+\frac{1}{2w}
-\Fwsmpa(w)
\right) e^{-w \hatt}
\nonumber \\
&~~~+\int_2^{\oo} dw \; 2 \left(\Fwla(w)
-\Fwdi(w)
-\Fwlapa(w)
\right)
e^{-w \hatt}
\ .
\end{align}

\section{Expansions about a finite point $\hatt=\hatt_0$}
First, we define the series removed of any leading factor
\begin{align}
\bar{f}_4^{(b;2;1)}(t)&=\tilde{f}_4^{(b;2;1)}(t)\; e^{2 \hatt}  \ ,\\
\bar{f}_4^{(b;2;2)}(t)&=\tilde{f}_4^{(b;2;2)}(t)\; e^{2 \hatt}
\sqrt{\hatt}/\ln \hatt\ , \\
\bar{f}_4^{(b;2;3)}(t)&=\tilde{f}_4^{(b;2;3)}(t)\; e^{2 \hatt}
\sqrt{\hatt}\ ,\\
\bar{f}_4^{(b;1;1)}(t)&=\tilde{f}_4^{(b;1;1)}(t)\; \hatt \ ,\\
\bar{f}_4^{(b;1;2)}(t)&=\tilde{f}_4^{(b;1;2)}(t)\; \hatt^2  \ .
\end{align}
Then, we expand about a finite point $\hatt=\hatt_0$ 
by substituting $t$ with 
$\hatt_0/(1+\uuu)^{1/2}$ 
in $\tilde{f}_4^{(b;1;x)}(t)$ 
and with  
$\hatt_0/(1+\uuu)$  
in $\tilde{f}_4^{(b;2;x)}(t)$,
and expanding in $\uuu$:
\begin{align}
\label{eq:f4v}
\facmm\bar{f}_4^{(b;1;1)}\left(\frac{\hatt_0}{\sqrt{1+\uuu}}\right)&=\sum_{n=0}^{\oo}
a^{(b;1;1)}_n \uuu^n \ ,\\
\facmm\bar{f}_4^{(b;1;2)}\left(\frac{\hatt_0}{\sqrt{1+\uuu}}\right)&=\sum_{n=0}^{\oo}
a^{(b;1;2)}_n \uuu^n \ ,\\
\facmm\bar{f}_4^{(b;2;1)}\left(\frac{\hatt_0}{1+\uuu}\right)&=\sum_{n=0}^{\oo}
a^{(b;2;1)}_n \uuu^n \ ,\\
\facmm\bar{f}_4^{(b;2;2)}\left(\frac{\hatt_0}{1+\uuu}\right)&=\sum_{n=0}^{\oo}
a^{(b;2;2)}_n \uuu^n \ ,\\
\facmm\bar{f}_4^{(b;2;3)}\left(\frac{\hatt_0}{1+\uuu}\right)&=\sum_{n=0}^{\oo}
a^{(b;2;3)}_n \uuu^n \ .
\end{align}
These particular substitutions $\hatt \to \uuu $ are chosen 
to improve the convergence of the series in $\uuu$
for $\hatt \to \oo$, corresponding to $\uuu\to -1$.

The coefficients $a_n^{(b;x;y)}$ can be obtained from the $w$-integral
representations
\eq{wintb11},
\eq{wintb12},
\eq{wintb21},
\eq{wintb22},
\eq{wintb23},
by expanding the integrands in $\uuu$ and integrating numerically term by term in $w$. 
Finally, $\tilde{f}_4^{(b)}(t)$ is worked out by summing up all the 6
contributions,
and the whole time-kernel $\tilde{f}_4(t)$ is recovered by adding 
also $\tilde{f}_4^{(a)}(t)$:
\begin{align}
\label{eq:f4fint}
\facmm
\tilde{f}_4(t) &=
\facmm
\tilde{f}_4^{(a)}(t)
+ \facmm \tilde{f}_4^{(b;1;3)}(t) 
\nonumber \\
&+\frac{1}{\hatt  } \sum_{n=0}^{\oo} a^{(b;1;1)}_n 
\left(\frac{\hatt_0^2}{\hatt^2}-1\right)^n 
\nonumber \\
&+\frac{1}{\hatt^2} \sum_{n=0}^{\oo} a^{(b;1;2)}_n
\left(\frac{\hatt_0^2}{\hatt^2}-1\right)^n 
\nonumber \\
&+e^{-2\hatt} \sum_{n=0}^{\oo} a^{(b;2;1)}_n
\left(\frac{\hatt_0}{\hatt}-1\right)^n 
\nonumber \\
&+\frac{e^{-2\hatt}}{\sqrt{\hatt}} \ln \hatt  \sum_{n=0}^{\oo} a^{(b;2;2)}_n
\left(\frac{\hatt_0}{\hatt}-1\right)^n 
\nonumber \\
&+\frac{e^{-2\hatt}}{\sqrt{\hatt}}         \sum_{n=0}^{\oo} a^{(b;2;3)}_n 
\left(\frac{\hatt_0}{\hatt}-1\right)^n  \ .
\end{align}

At this point we can use the expansions 
for small and for large $\hatt$ 
(\ref{eq:f4smallt}) and (\ref{eq:f4fint}) 
to get the values of 
$\tilde{f}_4(t)$ for any value of $\hatt$.

We choose a point of separation $\hatt=\hatt_s$ between the expansions.
In the region  $\hatt \le \hatt_s$ we compute 
$\tilde{f}_4(t)$ from the small-$t$ expansion~\eq{f4smallt}.
In the region $\hatt>\hatt_s$, we choose a suitable value 
of $\hatt_0$ and use~\eq{f4fint} to compute
$\tilde{f}_4(t)$.

The choice of the optimal $\hatt_s$,  $\hatt_0$ 
and the numbers of terms of the expansions 
depend on the level of precision required.
We choose  $\hatt_s=3.82$ and $\hatt_0=5$.
In Table~\ref{table:tabcoef} we list the coefficients
of the expansion (\ref{eq:f4fint}) up to $n={12}$.
If $\Delta\tilde{f}_4(\hatt)$ is the difference 
between the approximate value of $\tilde{f}_4(\hatt)$ obtained from the provided series expansions
and the value obtained by direct numerical integration,
the listed coefficients allow to obtain $\tilde{f}_4(\hatt)$
with an absolute error
$\facmm|\Delta\tilde{f}_4(\hatt)|<3\times10^{-8}$ for any value of
$\hatt \ge 0$, corresponding to a relative error 
$\left|\tfrac{\Delta\tilde{f}_4(\hatt)}{\tilde{f}_4(\hatt)}\right|<1.5\times 10^{-8}$.
Choosing an approximation to the correlation function $G(t)$ obtained
from the $R(s)$ data,
the relative error on the value of 
$a_{\mu}^{\rm HVP}(\textrm{NLO;4a})$ introduced by  using
our approximation to $\tilde{f}_4(\hatt)$
in the integral (\ref{eq:NLO4atimeintegral}) 
is 
$\Delta a_{\mu}^{\rm HVP}(\textrm{NLO;4a}) /a_{\mu}^{\rm HVP}(\textrm{NLO;4a}) \approx 10^{-13}$.

\begin{figure}
\begin{center}
\includegraphics[width=0.85\columnwidth]{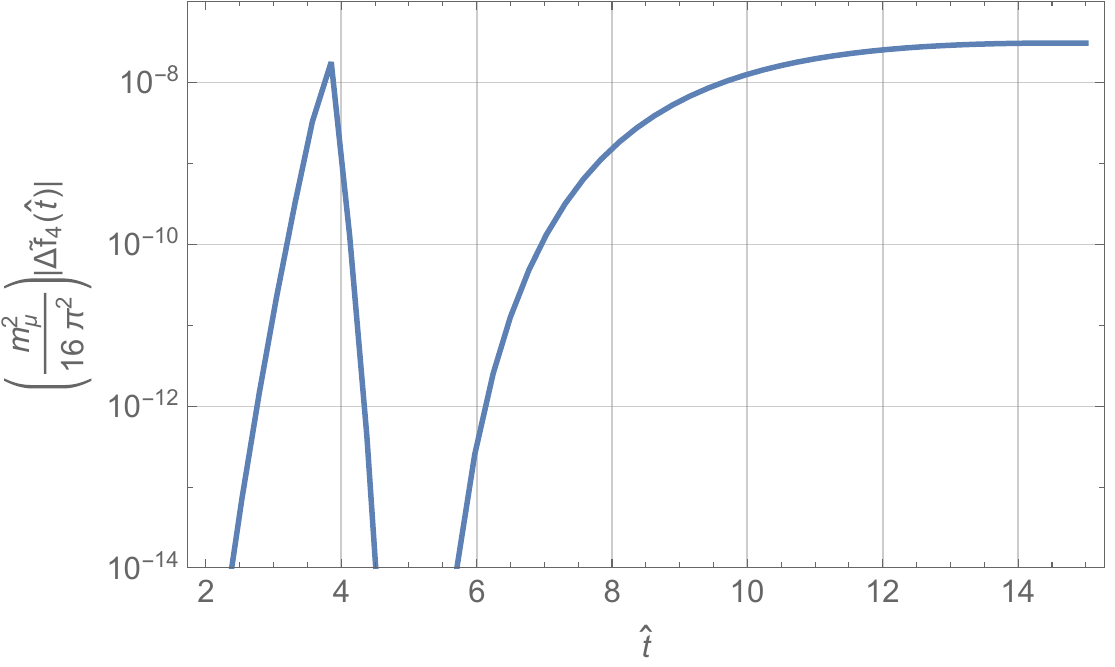}
\caption{Absolute value of the difference between the approximate
value of $\tilde{f}_4(t)$ obtained from the provided series expansions and the
value obtained by direct numerical integration.}
\label{fig:error}
\end{center}
\end{figure}

In Fig.\ref{fig:error} we show the behaviour of $|\Delta\tilde{f}_4(\hatt)|$. 
\section{Conclusions}
\label{Conclusions}

This paper provides series expansions of the time-momentum
representation of the NLO space-like kernel to be used in lattice QCD
computations of the NLO HVP contribution to the muon $g$-$2$.

After a derivation of the analytical time-momentum formula
for the HVP contribution at LO,
we analyzed in detail the components of the $\om-$integral
constituting the NLO time-momentum kernel.
One part was easily integrated analytically.
For the remaining part we were able to derive an expansion for small $\hatt$,
of which we computed the first 14 coefficients.

Next, we considered the expansion for large $\hatt$.
This expansion contains a part exponentially suppressed by a factor
$e^{-2 \hatt}$,  
and a part 
without any exponential factor, not present at LO.
We were able to derive the expansion for $\hatt \to \oo$ of these two parts,
which turned out to be  asymptotic,
and therefore of only limited use in numerical calculations.

These two main parts could be further divided into five different
subcontributions to the large-$\hatt$ expansion.
For each of these five parts we found a representation as an integral over
the imaginary frequency $w$.
Next, we expanded
these five
integrals
about a finite point $\hatt_0$
in powers of $(\hatt_0/\hatt-1)$.
These expansions turned out to be also converging  for 
$\hatt \to \oo$; we calculated the numerical values 
of the first 13 coefficients for $\hatt_0=5$.

The provided expansions  for 
small
and
large
$\hatt$
are sufficient to compute the time-kernel for every
value of $\hatt$ with an absolute error better than $3 \times 10^{-8}$,
which introduces a relative error on the value of
$a_{\mu}^{\rm HVP}(\textrm{NLO;4a}) $
of $O(10^{-13})$.

In conclusion, the results presented in this paper provide 
the NLO kernel in the time-momentum representation with a precision largely
sufficient for the lattice QCD determinations of
$a_{\mu}^{\rm HVP}$ at NLO accuracy.

{\em Acknowledgments} We would like to thank 
G.~Colangelo,
M.~Fael,
D.~Giusti,
M.~Hoferichter, 
T.~Teubner, 
G.~Venanzoni 
and H.~Wittig 
for useful discussions and correspondence. 
We are also grateful to all our MUonE colleagues for our stimulating 
collaboration.
S.~L. \ thanks the organizers of the ``Sixth Plenary Workshop of the Muon
$g$-$2$ Theory Initiative'' (Bern, 4-8 September 2023)
and the ``II Workshop of Muon Precision
Physics 2023'' (Liverpool, 7-10 November 2023) 
for providing support for attending the workshops.
S.~L.\ acknowledges support from the Italian Ministry of University
and Research (MUR) via the
PRIN 2022 project n.~20225X52RA --- MUS4GM2
funded by the European Union via the Next Generation EU package.
%
\bibliography{mybibfile}

\begin{table*}[]
\renewcommand{\arraystretch}{1.5}
\begin{center}
\begin{tabular}{|c|c|c|c|c|}
\hline
$\mathbf{n}$ & $\mathbf{a_n}$ & $\mathbf{b_n}$ & $\mathbf{c_n}$ &
$\mathbf{d_n}$ \\ \hline
   $4$    &  $\tfrac{317}{216}$  &   $-\frac{1}{3}$       &  $\frac{23}{18}$ &     $0$      \\ \hline
   $6$    &  $\frac{843829}{259200}$  &     $-\frac{371}{432}$     & $\frac{877}{1080}$ &      $\frac{19}{36}$     \\ \hline
   $8$    &   $\frac{412181237}{5292000}$ &    $-\frac{233}{48}$      & $-\frac{824603}{25200}$ &    $\frac{141}{20}$       \\ \hline
   $10$ &  $\frac{6272504689}{10584000}$  &    $-\frac{1165}{48}$      & $-\frac{460711}{1680}$ &      $\frac{961}{20}$     \\ \hline
   $12$ &  $\frac{404220031035193}{121022748000}$  &      $-\frac{42443}{378}$    & $-\frac{1359283213}{873180}$ &    $\frac{79342}{315}$       \\ \hline
   $14$ &  $\frac{14790819716039431}{890463974400}$  &     $-\frac{142931}{288}$     & $-\frac{4138386457}{540540}$ & $\frac{28243}{24}$          \\ \hline
   $16$ &  $\frac{38888413518277699}{503454631680}$  &    $-\frac{12895145}{6048}$      & $-\frac{489120278261}{13970880}$ &   $\frac{2605993}{504}$        \\ \hline
   $18$ &  $\frac{3950633085365067019}{11462583132000}$   &   $-\frac{116506871}{12960}$       & $-\frac{4589675124823}{29937600}$ &  $\frac{23642359}{1080}$         \\ \hline
   $20$ &  $\frac{364721869802634477577571}{243865691961091200}$   &    $-\frac{55559731}{1485}$      & $-\frac{37593205363634911}{57616158600}$ &      $\frac{44767436}{495}$     \\ \hline
   $22$ &  $\frac{77392239282793945882249}{12165635426630400}$  &     $-\frac{610873921}{3960}$     & $-\frac{26135521670035411}{9602693100}$ &     $\frac{121188929}{330}$      \\ \hline
   $24$ &  $\frac{27318770927965379913670522297}{1024872666654481444800}$  &    $-\frac{19509636989}{30888}$      & $-\frac{5138081420797732289}{459392837904}$ &     $\frac{3789385597}{2574}$      \\ \hline
   $26$ &  $\frac{449968490768168828714665100663}{4076198106012142110000}$  &   $-\frac{5618399257}{2184}$       & $-\frac{15810911801773817669}{348024877200}$ &   $\frac{151912159}{26}$        \\ \hline
   $28$ &  $\frac{251146293929498055156683549773}{554584776328182600000}$ &     $-\frac{678234361}{65}$     & $-\frac{3787066553671821473}{20715766500}$ &    $\frac{1495034796}{65}$       \\ \hline
   $30$ &  $\frac{100792117463017684643555224178269168501}{54680554570762463049907200000}$   &     $-\frac{2551294690547}{60480}$     & $-\frac{305996257628691658875533}{419236121304000}$ &    $\frac{64743309493}{720}$       \\ \hline
\end{tabular}
\end{center}
\caption{
Coefficients of the expansion of $\facmm \tilde{f}_4(t)$ up to $\hatt^{30}$, 
 see~\eq{f4smallt}
}.
\label{table:fsmallt}
\end{table*}

\begin{table*}[]
\renewcommand{\arraystretch}{1.0}
\begin{center}
\begin{tabular}{|c|c|c|c|c|c|}
\hline
 \boldmath{$n$ } &
 \boldmath{$a_n^{(b;1,1)}$} &
 \boldmath{$a_n^{(b;1,2)}$} & 
 \boldmath{$a_n^{(b;2,1)}$} & 
 \boldmath{$a_n^{(b;2,2)}$} & 
 \boldmath{$a_n^{(b;2,3)}$}\\  
\hline
0  & $ -1.4724671380$ & $  1.1589872337$ & $ -4.8942765691$ & $ 0.2973718753$ & $  2.1170734478$ \\
1  & $  0.1002442629$ & $ -0.0022459376$ & $ -2.9475017651$ & $ 0.4127862149$ & $  1.0364595246$ \\ 
2  & $  0.0021557710$ & $  0.0008279191$ & $ -0.5075497783$ & $ 0.1109534688$ & $  0.1101698869$ \\ 
3  & $  0.0001282655$ & $  0.0007999410$ & $  0.0115794503$ & $-0.0040980259$ & $  0.0167667530$ \\ 
4  & $ -0.0001467432$ & $ -0.0006094594$ & $ -0.0013940058$ & $ 0.0003899989$ & $ -0.0035236970$ \\
5  & $ 9.35581\times 10^{-6}$ & $  7.37693\times 10^{-6}$   & $ 0.0001421294$ & $ -0.0000133805$ & $ 0.0008586372 $ \\ 
6  & $  0.0000260037$ & $  0.0002711371$ & $ 7.67679\times10^{-6}$ & $ -0.00001764961$ & $  -0.0002257379 $ \\ 
7  & $ -0.0000189910$ & $ -0.0002551246$ & $-0.00001492424$ & $.000011742325$ & $  0.0000612688$ \\
8  & $ 6.93309\times 10^{-6}$ & $    0.0001291619$ & $8.61706\times 10^{-6}$ & $ -5.92454\times 10^{-6}$ & $-0.0000164422 $ \\
9  & $ 3.18779\times 10^{-7}$ & $   -0.0000121615$ & $-4.20065\times 10^{-6}$ & $ 2.78837\times 10^{-6}$ & $4.04750\times 10^{-6} $ \\ 
10 & $ -2.93399\times 10^{-6}$ & $  -0.0000553459$ & $1.95419\times 10^{-6}$ & $ -1.29025\times 10^{-6}$ & $-7.17744\times 10^{-7} $ \\
11 & $ 2.98580\times 10^{-6}$ &  $   0.0000760414$ & $-9.00478\times 10^{-7}$ & $ 5.98351\times 10^{-7}$ & $-7.67136\times 10^{-8} $ \\
12 & $ -2.08433\times 10^{-6}$ & $  -0.0000669985$ & $4.17032\times 10^{-7}$ & $ -2.80343\times 10^{-7}$ & $1.94188\times 10^{-7} $ \\
\hline
\end{tabular}
\end{center}
\caption{Coefficients of the expansions in $\uuu$
of $\facmm \tilde{f}_4(t)$  up to $\uuu^{12}$ with $\hatt_0=5$,
see~\eqq{f4v}{f4fint} 
}.
\label{table:tabcoef}
\end{table*}

\end{document}